\documentclass[12pt]{article}
\input epsf
\usepackage{amssymb}
\usepackage[dvips]{graphicx,psfrag}
\newcommand{\be}{\begin{equation}}
\newcommand{\ee}{\end{equation}}
\newcommand{\bea}{\begin{eqnarray}}
\newcommand{\eea}{\end{eqnarray}}
\newcommand{\lb}{\label}
\begin{document}
\begin{titlepage}
\title{On the equation of state of Dark Energy }
\author{David Polarski\thanks{email:polarski@lpta.univ-montp2.fr}~
and Andr\'e Ranquet\thanks{email:ranquet@lpta.univ-montp2.fr}\\
\hfill\\
Lab. de Physique Th\'eorique et Astroparticules, UMR 5207\\
Universit\'e Montpellier II,
34095 Montpellier Cedex 05, France}
\pagestyle{plain}
\date{\today}

\maketitle

\begin{abstract}
The formalism in order to obtain the Dark Energy equation of state is extended to non-flat 
universes and we consider the inequalities that must be satisfied by Phantom Dark Energy in this case. 
We show that due to a non-vanishing spatial curvature satisfying the observational bounds, 
the uncertainty on the determination of the Dark Energy equation of state parameter $w$, when it is 
taken constant, can be significant and that it is minimal for some redshift $z_{cr}\sim 3$. 
We consider the potential of future measurements of the gravitational waves emitted by binaries 
at high redshifts $z>z_{cr}$ to reduce this uncertainty. 
Results obtained here should also be relevant for a weakly varying equation of state with $w\approx -1$. 

\end{abstract}

\end{titlepage}

\section{Introduction}
There is growing observational evidence supporting the idea that the universe expansion departs 
radically from the standard picture \cite{P97}. Assuming a spatially flat isotropic 
universe, about two thirds of the critical energy density seems to be stored in a component 
called Dark Energy, able to produce an accelerated expansion. The precise nature of 
Dark Energy is still unknown and the subject of intense investigations and speculations \cite{SS00}. 
The main reason is that the simplest solution 
still fitting the data, a cosmological constant $\Lambda$, is not satisfactory as this implies an 
extraordinary fine tuning of the vacuum energy. On one hand it has to be unnaturally 
small -- about $123$ orders of magnitude smaller than its ``natural'' value -- and still not zero, 
on the other hand observations themselves may force us to consider Dark Energy with a varying 
equation of state \cite{ASSS04}. A popular and extensively studied possibility is a minimally coupled, 
slowly rolling, scalar field called Quintessence which would reproduce a mechanism quite similar to that 
of inflation \cite{RP88}. 
In these models, it is possible to obtain a negative pressure and accelerated 
expansion through an adequate balance of kinetic and potential energy. They satisfy the 
Null Energy Condition $\rho + p\ge 0$, which is saturated by  a cosmological constant $\lambda$, 
$\rho_{\Lambda} = -p_{\Lambda}$. However, a further surprise may come from the observational 
requirement to have Phantom Dark Energy with $\rho + p < 0$ \cite{C02}.
 
We have access to the expansion history of our universe at low redshifts $z$ using Supernovae 
data. By measuring their luminosity distances $d_L(z)$ we can reconstruct the Hubble diagram 
$H(z)$ with a differentiation procedure. To have access to the 
equation of state parameter $w(z)\equiv \frac{p(z)}{\rho(z)}$, one has to go through another 
differentiation prodecure, so the data do not allow to determine $w(z)$ sharply at the present time. 
On the other hand, this equation of state can give insight into the microscopic 
nature of Dark Energy. This is especially true if Dark Energy is of the Phantom type as it 
considerably restricts the possible candidates. Indeed, it is well known that the intensively 
studied quintessence models where the Dark Energy sector is some minimally coupled scalar field 
cannot account for an equation of state with $w<-1$. To account for this possibility, an 
extension of these models was originally suggested with the sign of the kinetic energy opposite to the 
conventional one \cite{C02}. Interestingly, we note that scalar-tensor theories of gravity provide 
Dark Energy models which allow for Phantom Dark Energy \cite{BEPS00},\cite{T02}, and where in addition 
the latter will be (weakly) clustered \cite{BEPS00}.
If observations force us to adopt Phantom Dark Energy, a whole class of models will be ruled 
out. It is further known that the future of our universe is dramatically different with 
Phantom Dark Energy \cite{CKW03}. Hence the special role played by the bordercase $w=-1$, also 
called the Phantom divide \cite{Hu04}.  

It is therefore of particular interest to investigate whether the observations will allow us to 
determine formally if Dark Energy (DE) is of the Phantom type or not. This is bound to become a 
crucial issue if observations gradually select a viable region in the vicinity of $w=-1$. 
So we would like to investigate the influence of priors concerning the geometry on the conclusions 
to be drawn from the observations. Usually, investigations assume a spatially flat 
universe. While such an assumption certainly has well founded theoretical motivations coming from 
the inflationary paradigm, it is clear that a small amount of curvature is still allowed by the 
Cosmic Microwave Background (CMB) data, even when combined with other data \cite{WMAP03},\cite{K05}.   
As we show in details, the assumption of spatial flatness i.e. $\Omega_{k,0}$ {\it exactly} zero, 
may lead us to confuse Phantom DE with usual DE. This kind of cosmic degeneracy or cosmic 
confusion is something recurrent in the interpretation of data. Clearly, the full problem requires a 
comprehensive statistical analysis involving all the cosmological parameters and general equations 
of state (see e.g.\cite{ASSS04},\cite{RCP05},\cite{Y05},\cite{CP01}). However, previous analyses 
assume spatial flatness and we want here to single out the degeneracy due to a non vanishing spatial 
curvature. As a first step, insight can be gained by considering a constant $w$ \cite{CP01} and 
working in the $(w,\Omega_{k,0})$ parameter plane. We show that measurements of the luminosity 
distances at high redshifts $z\ge 3$ can substantially restrict the degeneracy in the $(w,\Omega_{k,0})$ 
parameter plane in the neighbourhood of $(w=-1,\Omega_{k,0}=0)$. This shows the potential of high-$z$ 
$d_L$ measurements when combined with low-$z$ SNIa data by SNAP/JDEM (Supernova Acceleration Probe/
Joint Dark Energy Mission) \cite{SNAP} and we think this is a stimulating possibility worth to 
explore, especially in view of the future space mission LISA (Laser Interferometer Space Antenna) 
\cite{LISA} for the detection of low frequency gravitational waves.      
       
\section{The equation of state in non-flat universes}
In this Section we derive the basic equations related to the DE equation of state as well as 
the inequalities that can decide whether DE is of the Phantom type or not, when extended to 
non-flat universes. We assume our universe is isotropic with metric 
\be
ds^2 = dt^2 - a(t)~d\ell^2~, 
\ee
and its (expansion) dynamics is fully encoded in the time evolution of the scale factor $a(t)$ 
obeying the usual Friedmann-Lema\^\i tre-Robertson-Walker equations of General Relativity. 
The expansion of the universe can be probed through the measurement of the luminosity-distance 
$d_L(z)$ as a function of redshift $z$ (we take $c=1$)
\be
d_L(z) = (1+z)~H_0^{-1}~|\Omega_{k,0}|^{-\frac{1}{2}}~{\cal S}\left(|\Omega_{k,0}|^{\frac{1}{2}}~
\int_0^z \frac{dz'}{h(z')}\right)~.\lb{dLz}
\ee  
Here ${\cal S}(u)=\sin u$ for a closed universe, ${\cal S}(u)=\sinh u$ for an open universe 
while ${\cal S}$ is the identity for a flat universe. In eq.(\ref{dLz}), $h(z)$ stands for the 
(dimensionless) reduced Hubble parameter $h(z)\equiv \frac{H(z)}{H_0},~H_0\equiv H(z=0)$, and for 
small $z$ it is given by 
\be
h^2(z)= \Omega_{m,0} ~(1+z)^3 + \Omega_{DE,0} ~f(z) + \Omega_{k,0}~(1+z)^2~,\lb{hz}   
\ee
where $\Omega_{0}= \frac{\rho_{0}}{\rho_{cr,0}}$ for any component while 
$\Omega_{k,0}= -\frac{k}{a^2_0~H^2_0}$ and $\rho_{cr,0}=\frac{3H_0^2}{8\pi G}$.
The unknown function $f(z)$ expresses the evolution of Dark Energy with the expansion 
\be
f(z)\equiv \frac{\rho_{DE}(z)}{\rho_{DE,0}}~,\lb{fza} 
\ee   
and is directly related to the equation of state parameter $w(z)=\frac{p_{DE}(z)}{\rho_{DE}(z)}$. 
Indeed, the energy conservation equation valid for any isotropic perfect fluid
\be
\frac{d\rho}{dt} = -3 H (\rho + p)~,\lb{rhot}
\ee
where $t$ is the cosmological time, leads straightforewardly to
\be
f(z) = \exp \left[ 3\int_{0}^z dz'~\frac{1+w(z')}{1+z'}\right]~.\lb{fz}
\ee
For constant equation of state, we recover the well-known result $f(z)= (1+z)^{3(1+w)}$.
An important conclusion can be drawn from (\ref{fz}), or (\ref{rhot}), for Phantom DE ($w<-1$): 
its energy density decreases with increasing redshifts or increases in the course of time. 
This would give rise to a singularity in the future, the so called Big Rip. 

Though there is strong theoretical motivation from the inflationary paradigm in favour of a flat 
universe, we consider here spatially curved spaces in the conservative limits 
allowed by observations. 
The most stringent bound on the spatial curvature comes from the CMB data and these can be 
made even tighter in combinination with other data. They favour a spatially flat universe but 
some spatial curvature is still allowed. Actually the uncertainty on the 
spatial curvature increases with our lack of knowledge of DE, 
more specifically of its equation of state which is exactly what we dont know yet.

The observational signature for Phantom DE is that the following inequality be satisfied
\be
\frac{d h^2}{dz} < 3 ~\Omega_{m,0}~(1+z)^2 + 2 ~\Omega_{k,0}~(1+z)~.\lb{ineq}
\ee 
We see immediately from (\ref{ineq}) that the assumption of spatial flatness can lead to an 
erroneous conclusion: even though (\ref{ineq}) is satisfied assuming flatness, it could no longer 
be the case if the universe is closed. Similarly, if (\ref{ineq}) is not satisfied assuming 
flatness, it could still hold for an open universe. 
Assuming the following observational uncertainties
\bea
A &\le& \Omega_{m,0} \le C\\
-B &\le& \Omega_{k,0} \le D~,
\eea
where all constants $A,B,C,D$ are positive, we are {\it assured} DE is of the Phantom type if the following 
inequality is satisfied 
\be
\frac{d h^2}{dz} < 3 ~A~(1+z)^2 - 2 ~B~(1+z)~.\lb{ineqP}
\ee 
Note that $(\Omega_{k,0})_{min}\equiv -B$ is negative and corresponds to a closed universe.  
On the other hand we are assured DE is {\it not} of the phantom type if 
\be
\frac{d h^2}{dz}\ge 3 ~C~(1+z)^2 + 2 ~D~(1+z)~.\lb{ineqnP}
\ee 
For all other cases 
\be
3 ~A~(1+z)^2 - 2 ~B~(1+z) \le \frac{d h^2}{dz} < 3~C~(1+z)^2  + 2 ~D~(1+z)~,\lb{PnP}
\ee
uncertainties in the cosmological parameters allow for both types of DE. 

One should realize that the condition (\ref{ineq}) is local: Dark Energy is of the Phantom 
type for any $z$ where this condition is satisfied. Realization of 
(\ref{ineq}) even for one redshift value $z$ would constitute in itself a crucial result 
as it would discard all models that can not be of that type like quintessence models 
(minimally coupled scalar fields), and this would force us to consider alternative Phantom DE models.
On the other hand, the usefulness in this respect of the inequalities (\ref{ineq}) and 
(\ref{ineqP},\ref{ineqnP},\ref{PnP}) depends on the accuracy with which we can recover the 
quantity $h(z)$ from the data as a continuous function of redshift (see \cite{SASS05} for a 
recent discussion). Indeed, the use of these inequalities imply diffentiating twice the 
luminosity distance $d_L(z)$. A first differentiation will yield $h(z)$
\be
h^{-1}(z)= \left( \frac{D_L(z)}{1+z}\right)'~\left(1 + \left(\frac{D_L(z)}{1+z}\right)^2 
                                                      \Omega_{k,0} \right)^{-\frac{1}{2}}~,\lb{Hz}
\ee
where a prime denotes differentiation with respect to $z$ and we have introduced the dimensionless 
Hubblefree luminosity distance $D_L(z)\equiv H_0~d_L(z)$. The equation of state parameter $w(z)$ 
requires a second differentiation
\be
w(z) = \frac{\frac{1+z}{3} \frac{d h^2}{dz} - h^2 +\frac{1}{3}\Omega_{k,0}~(1+z)^2 }
                        {h^2 - \Omega_{m,0}~(1+z)^3 - \Omega_{k,0}~(1+z)^2 }~.\lb{wz}
\ee
We recognize from eq.(\ref{wz}) the expression for the DE pressure 
$p_{DE}(z)$ that can be recovered from the data
\be
p_{DE}(z) = (8\pi G)^{-1} \left((1+z)\frac{d H^2}{dz} - 3 H^2 +\Omega_{k,0}~H^2_0(1+z)^2\right)~.\lb{pDEz}
\ee
We see that $w_0\equiv w(z=0)$ is found from 
\be
w_0 = \frac{\frac{1+z}{3} \frac{d h^2}{dz}|_{z=0} - 1 +\frac{1}{3}\Omega_{k,0} }
                        {\Omega_{DE,0}}~,\lb{wz0}
\ee
where $\Omega_{DE,0}$ can be obtained from $\Omega_{m,0}$ {\it and} 
$\Omega_{k,0}$, namely 
\be
\Omega_{DE,0}=1-\Omega_{m,0}-\Omega_{k,0}~. 
\ee
Equations (\ref{Hz}-\ref{wz0}) 
generalize the corresponding flat space expressions, they apply to {\it arbitrary} equations of state 
and allow for a determination of $w(z)$ from the $d_L(z)$ data. 

\section{Behaviour in the parameter space $(w,\Omega_{k,0})$}

\begin{figure}[ht]
\begin{center}
\psfrag{k}[tl][br][2.0]{$\Omega _{k_0}$}
\psfrag{w}[][][2.0][-90]{$w$}
\includegraphics[angle=-90,width=.75\textwidth]{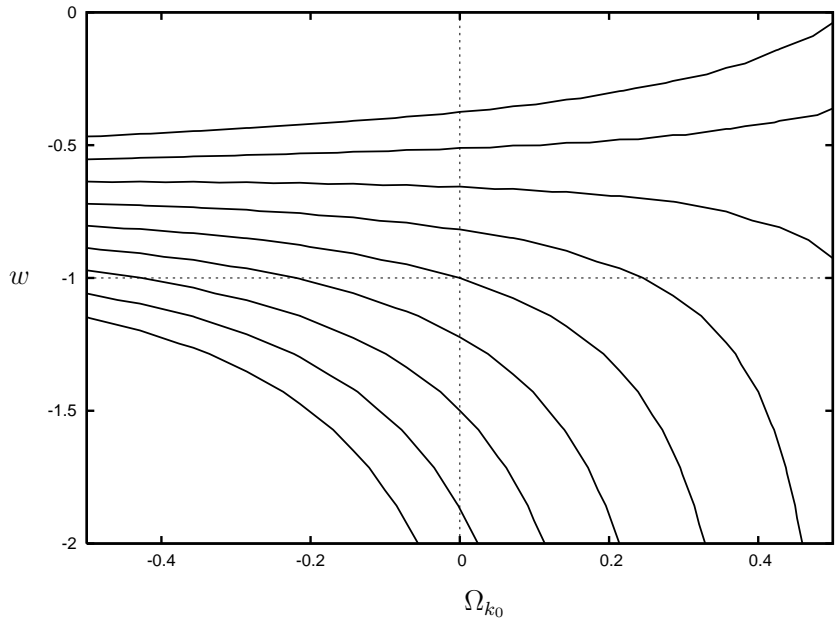}
\end{center}
\caption[]{Several iso-$d_L$ curves, curves corresponding to the same value of the luminosity distance $d_L$, 
are shown in the $(w,\Omega_{k,0})$ parameter plane for arbitrary but constant Dark Energy equations of state. 
The uncertainty in the universe curvature is essentially in the range $-0.15 \le \Omega_{k,0} \le 0.05$. 
It is seen that uncertainties on the value of $\Omega_{k,0}$ can induce significant uncertainties on the 
quantity $w$. This degeneracy is particularly interesting in the neighbourhood of the point 
$(w=-1,\Omega_{k,0}=0)$ because $w<-1$ (Phantom Dark Energy), $w=-1$ 
(cosmological constant $\Lambda$), and $w>-1$ all have dramatically different implications on the nature 
of Dark Energy.}
\lb{F1}
\end{figure}

\begin{figure}[ht]
\begin{center}
\psfrag{z}[tl][br][2.2]{$z_{cr}$}
\psfrag{w}[][][2.0][-90]{$w$}
\includegraphics[angle=-90,width=.75\textwidth]{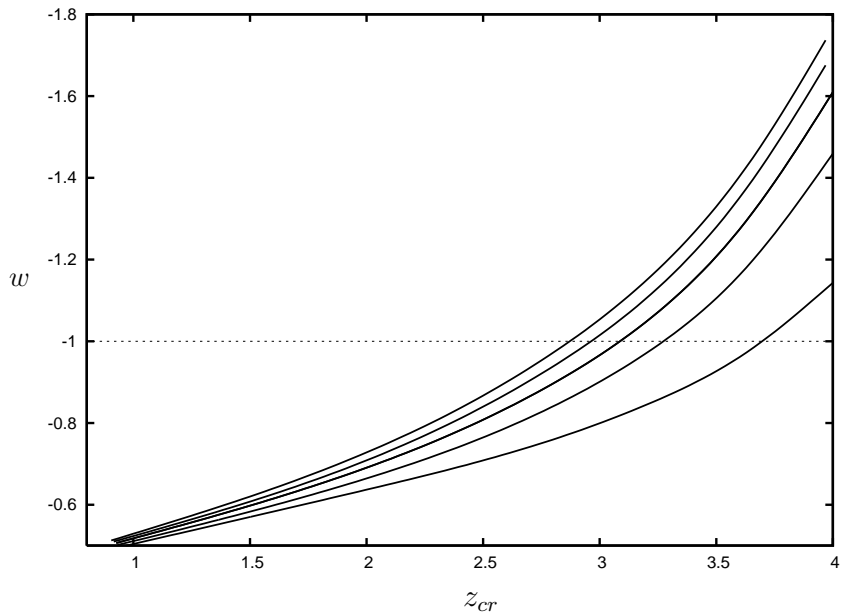}
\end{center}
\caption[]{Curves are displayed in the $(w,z_{cr})$ plane that give for arbitrary but constant $w$ the 
corresponding critical redshift value $z_{cr}$ for which 
$\frac{\partial d_L(z)}{\partial \Omega_{k,0}}(z=z_{cr},\Omega_{k,0}=0)=0$. Each curve corresponds to 
a fixed value of the cosmological parameter $\Omega_{m,0}$ and we have from top to bottom 
$\Omega_{m,0}=0.5, 0.4, 0.3, 0.2, 0.1$. For the value $w=-1$, the coresponding redshift interval is 
$2.8\le z_{cr}\le 3.7$ with $z_{cr}=3.1$ for the fiducial value $\Omega_{m,0}=0.3$. At $z=z_{cr}$, 
the degeneracy is maximal in the $\Omega_{k,0}$ direction, but minimal in $w$.} 
\lb{F2}
\end{figure}

We want to investigate the amount of cosmic confusion in the region around $w=-1$ that can arise due 
to a non vanishing curvature of the universe. As said in the Introduction, we will work in the two 
dimensional parameter space $(w,\Omega_{k,0})$ for fixed $\Omega_{m,0}$ and arbitrary constant $w$. 
The observational bounds on the geometry of the universe coming from the WMAP (Wilkinson Microwave 
Anisotropy Probe) data \cite{WMAP03}, possibly combined with other data, are the most stringent ones. 
They favour a marginally closed, nevertheless close to flatness, universe which is reassuring for the 
inflationary paradigm.     
The bounds extracted from the data depend on the various priors about the underlying model 
universe and most crucially on the nature of DE. We certainly don't want to restrict ourselves to 
$w=-1$ and uncertainties in the DE equation of state will inevitably relax the observational bounds. 
We will take the following interval of interest \cite{K05} 
\be
-0.15 \le \Omega_{k,0} \le 0.05~.
\ee 
On the other hand, the data allow $w=-1$ as a good fit to the SNIa data. So it is clear 
that the interesting region is centered around $(w=-1,\Omega_{k,0}=0)$ and we can expect 
future data will continue to single out this region .

We consider first luminosity-distances $d_L$ in our two dimensional parameter space, measured at 
{\it fixed} redshift $z$. 
It is clear from the iso-$d_L$ curves, curves having the same $d_L$ value, displayed on Figure 1 for 
fixed $z=1.5$ how cosmic confusion can arise: observations interpreted as a pure cosmological constant 
$\Lambda$ {\it assuming} a flat universe can equally well be interpreted as either DE with $w>-1$, 
resp. Phantom DE ($w<-1$), if one takes a closed, resp. open, universe. 

The crucial point is how this evolves with increasing redshift. 
In the region of parameter space of interest, we find that this behaviour changes at some critical redshift 
$z_{cr}\sim 3$. Indeed at $z=z_{cr}$, there is essentially no degeneracy in $w$ 
while we have a maximal degeneracy with respect to curvature
\be
\frac{\partial d_L(z=z_{cr})}{\partial \Omega_{k,0}}|_{\Omega_{k,0}=0} = 0~.
\ee     
The value of $z_{cr}$ depends of course both on the values of $\Omega_{m,0}$ and $w$. 
We have checked that a corresponding behaviour takes place when we consider different values 
$0.1\le \Omega_{m,0}\le 0.5$ and different $w$, as is summarized in Figure 2 where we can read the 
critical redshift that corresponds to various values $\Omega_{m,0}$ and $w$. 
So, for given $w$ of interest, our analysis can be repeated for any $\Omega_{m,0}$ as its value will be 
determined with increasing accuracy.

The degeneracy in the $(w,\Omega_{k,0})$ parameter plane is opposite for higher redshifts $z>z_{cr}$: 
the iso-$d_L$ curve through the point $(w=-1,\Omega_{k,0}=0)$ fits also a closed universe with Phantom 
DE as well as an open universe with $w>-1$. One could use this property in order to constrain the 
degeneracy in $w$ as can be seen from Figures 3, 4. 
It could also be used in order to constrain substantially the geometry of our universe. However 
one should remember that our analysis is done for {\it non-varying} equations of state and that the 
precision with which we can recover the luminosity distance at $z>z_{cr}$ is a crucial issue that 
cannot yet be settled. 
On the other hand, results found here will essentially hold if Dark Energy has a weakly varying 
equation of state, and they would become particularly interesting if $w\approx -1$. 

\begin{figure}[ht]
\begin{center}
\psfrag{k}[tl][br][2.0]{$\Omega _{k_0}$}
\psfrag{w}[][][2.0][-90]{$w$}
\includegraphics[angle=-90,width=.75\textwidth]{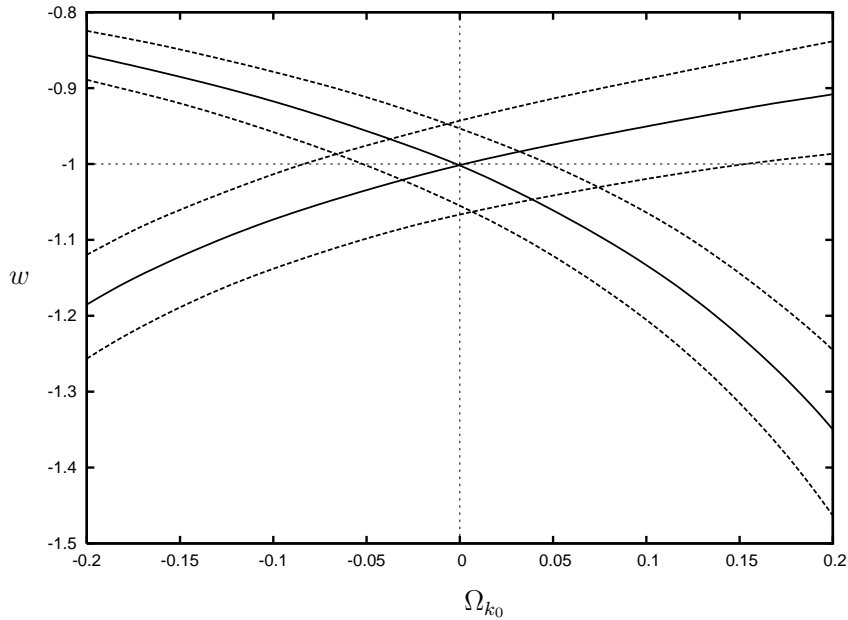}
\end{center}
\caption[]{Two iso-$d_L$ curves (solid) are shown for $\Omega_{m,0}=0.3$ in the $(w,\Omega_{k,0})$ 
plane that pass through the point $(w=-1,\Omega_{k,0}=0)$, the first (upper solid curve on the left) 
correspond to $z=1$ while the second solid curve corresponds to $z=4$. The corresponding $1\%$ errors 
(dashed lines) are also shown for each curve. It is seen that if measurements at both redshifts fit 
the couple $(w=-1,\Omega_{k,0}=0)$, the degeneracy in the value of $w$ that comes from the possibility 
to have non-flat universes could be substantially restricted as well as the spatial curvature.}
\lb{F3}
\end{figure}

\section{High $z$ measurements using gravitational waves}
High redshifts measurements of the luminosity distance $d_L$ could be made using 
the emission of gravitational waves (GW) by inspiralling black holes binaries at high redshifts $z$ 
\cite{B04}. 
The flux $F$, the amount of energy per unit of time and per unit of area in gravitational waves 
produced by such binaries that we measure, is given by (we put back the velocity of 
light $c$) 
\be
F = \frac{c^3}{16 \pi G} ( {\dot h}_+^2 + {\dot h}_{\times}^2 )~,\lb{FGW}
\ee
where $h_+$, $h_{\times}$ are the two polarization states. 
Formally we can express $dL$, the energy emitted by the system per unit of time (in the binary's 
restframe) in the solid angle $d\Omega$, in the following way
\be
dL = \frac{c^3}{16 \pi G} ~( {\dot h}_+^2 + {\dot h}_{\times}^2 )~d_L^2 ~d\Omega~,
\ee
so that, with $a_0$ the present value of the scale factor and $r_1$ the 
coordinate distance to the binary, the measured flux $F$ becomes
\be
F = \frac{dL}{dA ~(1+z)^2} = \frac{dL}{a_0^2~r_1^2 ~d\Omega ~(1+z)^2} = \frac{1}{d_L^2} ~\frac{dL}{d\Omega}~,
\ee
and (\ref{FGW}) is recovered. 
The polarisation states $h_+$, $h_{\times}$ are given by 
\bea
h_+ &=& \frac{2 ~G^{\frac{5}{3}}}{c^4} ~\frac{{\cal M}^{\frac{5}{3}}}{d_L}~(\pi f(t))^{\frac{2}{3}}~
                           [1+({\vec L}.{\vec n})^2]~\cos 2 \Phi(t) \\  
h_{\times} &=& \frac{2 ~G^{\frac{5}{3}}}{c^4} ~\frac{{\cal M}^{\frac{5}{3}}}{d_L}~(\pi f(t))^{\frac{2}{3}}~
                           (2~{\vec L}.{\vec n})~\sin 2 \Phi(t)~.
\eea
The unit vector ${\vec n}$ defines the binary's position on the sky while ${\vec L}$ points along the 
binary's orbital angular momentum and hence defines its orientation. The important quantity ${\cal M}$ 
is the chirp mass given by 
\be
{\cal M} = (1+z)~\frac{(m_1~m_2)^{\frac{3}{5}}}{(m_1+m_2)^{\frac{1}{5}}}~,\lb{Mc}
\ee 
where $m_{1,2}$ are the two masses. As can be seen from (\ref{Mc}), ${\cal M}$ contains the factor $1+z$. 
The data allow us to determine ${\cal M}$ accurately but without independent knowledge of the masses 
$m_{1,2}$, we cannot deduce the redshift $z$ from the data. 
Finally the orbital phase $\Phi(t)$ is related to the wave frequency $f(t)$, $\frac{d\Phi}{dt}= \pi~f(t)$ 
while ${\dot f(t)}$ depends on the chirp mass ${\cal M}$, ${\dot f} \sim f^{\frac{11}{3}} ~{\cal M}^{\frac{5}{3}}$.
%
%
Extracting ${\dot f}$, or ${\cal M}$, from the data allows us to find $d_L$, for example we can use 
the ratio $\frac{h}{\dot f}$ 
%
%
\be
d_L \sim \frac{\dot f}{h_{\times,+}}~\frac{g_{\times,+}}{f^3}~,
\ee 
where $g_{\times,+}$ encodes the dependence on the binary's orbital phase, position and inclination 
and can be determined from the data.
To summarize, we can extract from the GW data the quantity $d_L$, however {\it not} the redshift $z$ 
that corresponds to it. 

The spatial interferometer LISA for the detection of gravitational waves around frequencies 
$\nu\sim (10^{-3}-10^{-4})$ Hz that will operate in the future could possibly make such high $z$ measurements. 
For example the gravitational waves emitted by inspiralling binary black holes with large masses 
$m_{1,2}\sim 10^5~M_{\odot}$ at redshifts $z>z_{cr}=3 - 4$, and even at higher redshifts, 
could be detected by LISA and this detection would yield a measurement of the distance $d_L$ 
of the source but not of its redshift. If in addition, some electromagnetic counterpart to the 
GW emission can be identified, then the situation changes dramatically. By determining the position 
of the source, the relative error on the luminosity distance could even drop below the 
$1 \%$ level, $\frac{\delta d_L}{d_L} \sim 1 \%$ \cite{HH05}.
Note that these black hole binaries could constitute standard candles (``sirens'') 
complementary to the SNIa. 

In this way, new $d_L$ points at high redshifts could be obtained with high accuracy. 
Clearly this could considerably restrict the possible degeneracy in the 
$(w,\Omega_{k,0})$ plane and it is this aspect we want to stress here. 
We will assume for reference that measurements of $d_L(z)$ at $z\sim 4$ are made at the $1\%$ 
precision level. 

\begin{figure}[ht]
\begin{center}
\psfrag{k}[tl][br][2.0]{$\Omega _{k_0}$}
\psfrag{w}[][][2.0][-90]{$w$}
\includegraphics[angle=-90,width=.75\textwidth]{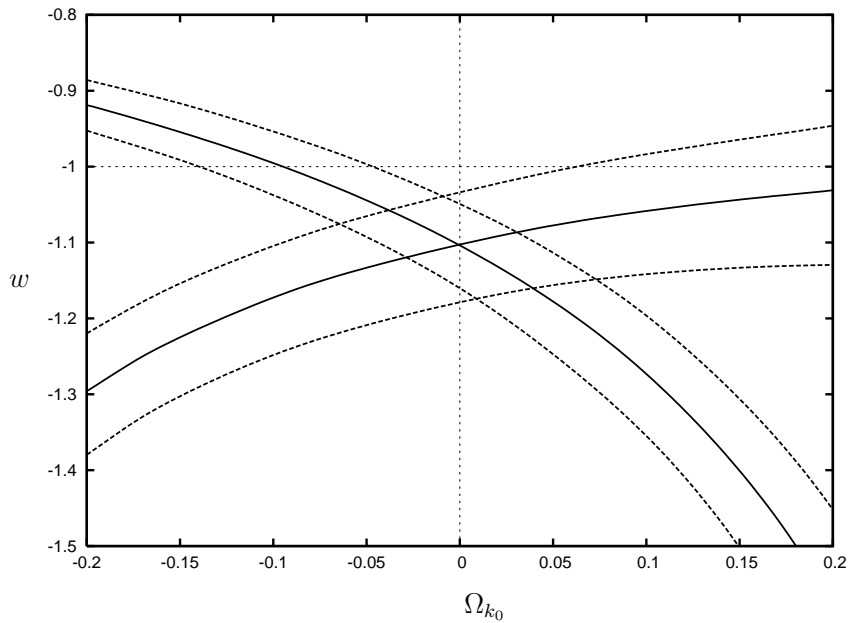}
\end{center}
\caption[]{As in Figure 3, two iso-$d_L$ curves (solid) are shown, but this time they fit at both 
redshifts $z=1$ (upper solid curve on the left) and $z=4$ the values $(w=-1.1,\Omega_{k,0}=0)$. 
With $1\%$ errors at both redshifts, it is seen that the equation of state parameter $w$ is constrained 
to satisfy $w<-1$.}
\lb{F4}
\end{figure}

Let us assume for example that high redshifts, as well as low redshifts luminosity distance 
measurements fit a universe with $w=-1$ and $\Omega_{k,0}=0$. 
Then, as can be seen from Figure 3, the degeneracy in $w$ would become minimal, essentially that 
of measurements made at $z=z_{cr}$. 
It would be natural to interpret this as support for a flat $\Lambda$ dominated universe. 
A similar reasoning would apply if observations, assuming a flat universe, fit a 
constant $w$ different from $-1$: if $d_L$ measurements at high and low redshifts are obtained that fit 
a flat universe with Phantom DE, the uncertainty on the equation of state could be small enough so as 
to provide strong evidence for Phantom DE, again assuming a constant $w$, as we show with an example 
in Figure 4. We see also from Figure 4 that we must have $w\lesssim -1.06$ in order to characterize 
phantom Dark Energy unambiguously.

It would be interesting to investigate how a varying equation of state would modify these 
conclusions and we leave this for further investigation. Clearly, such an investigation will 
become more effective once we have enough information about the time-varying equation of state 
that could be extrapolated to redshifts $z\ge z_{cr}$. It is clear however that our 
results will essentially hold for a weakly varying equation of state. In this case accurate 
results can only be obtained for given equation of state. 
In conclusion, we have shown how the complementarity of low redshift SNIa data with high redshift 
measurements by LISA could provide us with useful information on the nature of Dark Energy. 
We feel this complementarity is an interesting direction to explore in particular if one relaxes the 
assumption of spatial flatness.

\section*{Acknowledgments}
It is a pleasure to thank L. Blanchet for illuminating discussions and A. Barrau for stimulating 
comments.


\begin{thebibliography}{99}


\bibitem{P97} S.J. Perlmutter et al., Ap. J. {\bf 483} 565 (1997), Nature {\bf 391} 51 (1998);
 A.G. Riess et al., Astron. J. {\bf 116} 1009 (1998)
 R. A. Knop {\it et al}, Astroph. J.{\bf 598}, 102 (2003);
 J.L. Tonry et al., Ap. J. {\bf 594} 1 (2003);
 A. G. Riess {\it et al}, Astroph. J.{\bf 607}, 665 (2004);
 U. Seljak {\it et al}, {\tt astro-ph/0407372}
\bibitem{SS00} V. Sahni, A. A. Starobinsky, Int. J. Mod. Phys. D{\bf 9}, 373 (2000); 
 V. Sahni, {\tt astro-ph/0502032}; 
 T. Padmanabhan, Phys. Rep. {\bf 380}, 235 (2003)
\bibitem{ASSS04}  U. Alam, V. Sahni, T. D. Saini, A. A. Starobinsky, Mon. Not. Roy. Astron. Soc. {\bf 354}, 275 (2004)
\bibitem{RP88} B. Ratra and P.J.E. Peebles, Phys. Rev. D{\bf 37} 3406 (1988);
 C. Wetterich, Nucl. Phys. B{\bf 302} 668 (1988);
 P.G. Ferreira and M. Joyce, Phys. Rev. Lett. {\bf 79} 4740 (1997);
 R.R. Caldwell, R. Dave and P.J. Steinhardt, Phys. Rev. Lett. {\bf 80} 1582 (1998)
\bibitem{C02} R. Caldwell, Phys. Lett. B {\bf 545}, 23 (2002)
\bibitem{CKW03} R. Caldwell, M. Kamionkowski, N. N. Weinberg, Phys. Rev. Lett. {\bf 91} 043503 (2003);
 M. Dabrowski, T. Stachowiak, M. Szydlowski, Phys. Rev. D{\bf 68}, 103519 (2003)
\bibitem{Hu04} W. Hu, {\tt astro-ph/0410680} 
\bibitem{BEPS00} B. Boisseau, G. Esposito-Far\`ese, D. Polarski and A.A. Starobinsky, 
 Phys. Rev. Lett. {\bf 85} 2236 (2000)
\bibitem{T02} D. Torres, Phys. Rev. D{\bf 66}, 043522 (2002)
\bibitem{WMAP03} D. N. Spergel {\it et al}, Ap. J. S. 148 (2003) 175
\bibitem{K05} L. Knox, {\tt astro-ph/0503405}
\bibitem{RCP05} T. Roy Choudhury, T. Padmanabhan, Astron. Astrophys. {\bf 429}, 807 (2005)
\bibitem{Y05} Ch. Yeche, A. Ealet, A. Refregier, C. Tao, A. Tilquin, J.-M. Virey, D. Yvon, {\tt astro-ph/0507170};
 H. K. Jassal, J.S. Bagla, T. Padmanabhan, {\tt astro-ph/0506748}; 
 P.-S. Corasaniti, T. Giannantonio, A. Melchiorri, Phys. Rev. D{\bf 71}, 123521 (2005);
 D. Huterer, M. Takada, G. Bernstein, B. Jain, {\tt astro-ph/0506030};
 K. Glazebrook, C. Blake, {\tt astro-ph/0505608};
 C. Espana-Bonet, P. Ruiz-Lapuente, {\tt hep-ph/0503210};
 R. Lazkoz, S. Nesseris, L. Perivolaropoulos, {\tt astro-ph/0503230};
 M. Jarvis, B. Jain, G. Bernstein, D. Dolney, {\tt astro-ph/0502243};
 A. Upadhye, M. Ishak, Paul J. Steinhardt, {\tt astro-ph/0411803};
 D. Rapetti, S. W. Allen, J. Weller, Mon. Not. Roy. Astron. Soc. {\bf 360}, 555 (2005);
 B. A. Bassett, P.-S. Corasaniti, M. Kunz, Astrophys.J.{\bf 617}, L1 (2004); 
 V. B. Johri, {\tt astro-ph/0409161};
 D. A. Dicus, W. W. Repko, Phys. Rev. D{\bf 70}, 083527 (2004);
 D. Jain, J.S. Alcaniz, A. Dev, {\tt astro-ph/0409431};
 Bo Feng, X.-L. Wang, X.-M. Zhang, Phys. Lett. B{\bf 607}, 35 (2005);
 Y.-G. Gong, Int. J. Mod. Phys. D{\bf 14}, 599 (2005); 
 E. V. Linder, Phys. Rev. Lett. {\bf 90}, 091301 (2003);
 I. Maor, R. Brustein, J. McMahon, Paul J. Steinhardt, Phys.Rev.D {\bf 65}, 123003 (2002)
\bibitem{CP01} M. Chevallier, D. Polarski, Int. J. Mod. Phys. D{\bf 10}, 213 (2001) 
\bibitem{SNAP} http://snap.lbl.gov/
\bibitem{LISA} http://lisa.jpl.nasa.gov/ 
\bibitem{SASS05} A. Shafieloo, U. Alam, V. Sahni, A. A. Starobinsky, {\tt astro-ph/0505329}
\bibitem{B04} L. Blanchet, T. Damour, G. Esposito-Far\`ese, B. Iyer, Phys. Rev. Lett. {\bf 93} 091101 (2004)
\bibitem{HH05} D. E. Holz, S. A. Hughes, {\tt astro-ph/0504616};  
 B. Kocsis, Z. Frei, Z. Haiman, K. Menou, {\tt astro-ph/0505394}

\end{thebibliography}
\end{document}